\documentclass[aps,prd,10pt,superscriptaddress,twocolumn,nofootinbib]{revtex4-1}
\usepackage{graphicx, epsfig}
\usepackage{amsmath,amssymb,amsfonts,dsfont,mathrsfs,amsthm,mathtools}
\usepackage{bm}
\usepackage{color}
\usepackage{physics}
\usepackage[usenames]{xcolor}
\usepackage{hyperref}
\usepackage{siunitx}
\hypersetup{colorlinks=true,urlcolor=blue,linkcolor=blue,citecolor=blue,filecolor=blue}
\usepackage[normalem]{ulem}
\usepackage{array}
\usepackage{hyperref}
\usepackage{booktabs}
\usepackage{cancel}
\usepackage{blindtext}
\usepackage{tikzsymbols}
\usepackage[paperwidth=625pt,top=80pt,right=60pt,bottom=80pt,left=60pt,headheight=50pt]{geometry}
\usepackage{lipsum}
\usepackage{orcidlink}

\newcommand{\Lag}{\mathscr{L}}

\begin{document}

\title{Hadronic lensing}
\begin{abstract}
We introduce an analytic approach to study gravitational lensing in the presence of a distribution of hadrons. The situation is analogous to the propagation of photons in a medium with a nontrivial Cooper-pair condensate, where the photon acquires an effective mass term that may depend on the coordinates if the condensate is not homogeneous. As a result, photons generally do not follow null geodesics in the hadronic medium. In this setup, hadrons are described by the nonlinear sigma model minimally coupled to Maxwell theory. The modified Raychaudhuri equation, including hadronic corrections, is derived, along with the integral curves of probe photons in the eikonal approximation. These results are consistent with the theory of gravitational lensing in plasma media, with the advantage that transport properties, such as the refractive index, can be expressed analytically in terms of the hadronic density without assuming a phenomenological modeling thereof. As an example, we study the hadronic lensing produced by an analytic black hole sourced by superfluid pionic vortices, and we obtain the hadronic correction to the deflection angle in the weak-field limit. 
\end{abstract}

\author{Fabrizio
\surname{Canfora}\,\orcidlink{0000-0002-4661-9875}
}
\email{fabrizio.canfora@uss.cl}
\affiliation{Centro de Estudios Científicos (CECs), Casilla 1469, Valdivia, Chile}
\affiliation{Facultad de Ingenier\'ia, Universidad San Sebastian \\ General Lagos 1163, Valdivia 5110683, Chile}

\author{Crist\'obal \surname{Corral}\,\orcidlink{0000-0002-4503-4154}
}
\email{cristobal.corral@uai.cl}
\affiliation{Departamento de Ciencias, Facultad de Artes Liberales, Universidad Adolfo Ib\'añez, Avenida Padre Hurtado 750, 2562340, Viña del Mar, Chile}

\author{Borja \surname{Diez}\,\orcidlink{0009-0004-3805-4036}
}
\email{borja.diez@cinvestav.mx}
\affiliation{Departamento de F\'{\i}sica, Cinvestav, Av.~IPN 2508, 07360, CDMX, M\'exico}

\maketitle


\section{Introduction}
Gravitational lensing is one of the most robust observational predictions of General Relativity (GR). It has become a fundamental tool for studying compact astrophysical objects, mapping the large-scale distribution of matter, and testing theories of gravity in both weak and strong-field regimes~\cite{Kaiser:1992ps,Bartelmann:1999yn,Hoekstra:2013via,mandelbaum2014galaxy,Giocoli:2013tga}. Additionally, it also plays a central role in the analysis of the cosmic microwave background (CMB) and related cosmological observables~\cite{Lewis:2006fu,Nguyen:2017zqu,Marozzi:2016und,Peloton:2016kbw,
Fabbian:2017wfp,Pratten:2016dsm,Hagstotz:2014qea,Bonvin:2015uha,
Marozzi:2016uob,Petri:2016qya,Schaefer:2005up,Cooray:2002mj,Marozzi:2016qxl}. In the context of neutron stars, recent observations of radio and X-ray pulse profiles, together with polarization measurements obtained by NuSTAR, NICER, and IXPE~\cite{Yu:2024lzc,Bobrikova:2024soh,Benacek:2025wgp}, have enabled increasingly precise determinations of mass-radius relations~\cite{NuSTAR:2013yza,Vinciguerra:2023qxq,Ursini:2023wjf}. These measurements, in turn, can be used to impose constraints on the internal composition and equation of state of dense nuclear matter~\cite{Bogdanov:2019qjb,Raaijmakers:2019dks}, allowing one to perform a phenomenological analysis of light propagation in plasma environments near compact objects~\cite{Bisnovatyi-Kogan:2010flt,Tsupko:2013cqa,Er:2013efa,Morozova:2013uyv,Rogers:2015dla,Kobialko:2025sls,Steiner:2024aum}. In neutron stars, for instance, electromagnetic fields interact nontrivially with strongly coupled nuclear matter. As a result, photon propagation is no longer described solely in terms of null geodesics of the background metric. In such a case, light rays propagate according to an effective optical geometry, and their trajectories deviate from null geodesics of the gravitational metric. Accounting for these effects is essential for a consistent description of light propagation in strongly interacting environments.

A useful physical analogy arises from superconductivity. In ordinary conductors, electromagnetic fields penetrate the material with only mild qualitative modifications. In contrast, superconductors exhibit the well-known Meissner effect~\cite{Meissner:1933ela}, whereby magnetic fields are expelled due to the collective behavior of the condensed phase~\cite{kozhevnikov2025electrodynamics}. Ignoring this phenomenon would lead to an incomplete description of electromagnetic propagation inside superconducting media. Similarly, analyzing gravitational lensing while neglecting the role of self-gravitating hadronic matter overlooks essential physical effects arising from the source's microscopic structure. Just as the Meissner effect represents a macroscopic manifestation of underlying quantum interactions, the presence of strongly coupled hadronic matter may significantly alter the effective propagation of light through spacetime.

Motivated by these considerations, here we explore gravitational lensing in spacetimes with self-gravitating hadronic matter. Our aim is to determine how their backreaction modifies the effective propagation of light and to identify observable deviations from the standard null-geodesic approximation. Concretely, we focus on the nonlinear sigma model (NLSM) with global $SU(2)$ isospin symmetry coupled to Maxwell theory, which corresponds to the leading-order expansion of chiral perturbation theory in the pion decay constant. This setup provides one of the most relevant low-energy effective descriptions of hadronic physics~\cite{Manton:2004tk,Shifman:2012zz,Shuryak:2021vnj}. In this theory, the electromagnetic field couples nontrivially to the pionic sector. As a consequence, in backgrounds dressed with hadronic matter, electromagnetic waves no longer propagate along null geodesics of the spacetime metric. This can be seen explicitly by taking the eikonal limit of the Maxwell equations, which reveals modifications to the dispersion relation. Furthermore, these corrections modify the Raychaudhuri equation~\cite{Raychaudhuri:1953yv} governing congruences, reflecting a departure from standard geodesic behavior. This framework, therefore, provides a natural arena for investigating how strongly interacting matter influences photon's propagation and assessing the observational relevance of such effects in compact astrophysical systems.
 
This work is organized as follows. In Sec.~\ref{sec:theory}, we present our theoretical setup, namely General Relativity with cosmological constant coupled to the nonlinear sigma model and fix our notation. Then, in Sec.~\ref{sec:eikonal}, we derive the appropriate eikonal limit of Maxwell fields propagating on a background with self-gravitating hadronic matter, and obtain both the modified Raychaudhuri equation and the integral curve that describes the photon trajectory in such backgrounds. In Sec.~\ref{sec:lensing-bumby-BH} we consider, as a concrete example, a black hole sourced with a nontrivial distribution of superfluid pions, recently found in Ref.~\cite{Canfora:2026col}. Then, we show how the gravitational lensing of light rays is modified by the Hadronic matter. Finally, in Sec.~\ref{sec:conclusions}, we summarize our results and present our conclusions.

\section{The theory}\label{sec:theory}
The dynamics of the NLSM coupled to General Relativity is dictated by the action principle~\cite{Scherer:2002tk}
\begin{equation}\label{action}
\begin{split}
	I[g,U]&=\int_\mathcal{M}\dd^4x\sqrt{-g}\left[\frac{R-2\Lambda}{2\kappa}+\frac{K}{4}\Tr(L_\mu L^\mu)\right]\,,
\end{split}
\end{equation}
where $\kappa=8\pi G$ is the gravitational constant, $\Lambda$ is the cosmological constant, while $K$ is a positive coupling fixed by experimental data. The next-to-leading corrections in the chiral perturbation theory ($\chi$PT) expansion can be systematically included; we will return to them in a future publication. Additionally, $U(x)\in SU(2)$ represents the pionic field, and we have defined
\begin{equation}
	L_\mu=U^{-1}\nabla_\mu U=L_\mu^{j}t_j\,,
\end{equation}
with $t_j=i\sigma_j$ being the (anti-)Hermitian generators of $SU(2)$ and $\sigma_j$ are the Pauli matrices. The pionic field can be parametrized as
\begin{align}\label{U-parametrization}
	U(x)=\cos\alpha \,\mathbb{I}_{2\times 2} +\sin\alpha\,n^{j}t_j\,,
\end{align}
where the vector $n^{i}=(\sin\Theta\cos\Phi,\sin\Theta\sin\Phi,\cos\Theta)$ satisfies $n^{j}n_j=1$, while the scalar functions $\alpha=\alpha(x)$, $\Theta=\Theta(x)$, and $\Phi=\Phi(x)$ represent the three dynamical degrees of freedom of the pionic field.

The field equations for the metric and pions obtained from the action~\eqref{action} are given by
\begin{subequations}\label{eom-g-U}
	\begin{align}
		R_{\mu\nu}-\frac{1}{2}g_{\mu\nu}R+\Lambda g_{\mu\nu}&=\kappa T_{\mu\nu}\label{eom-g}\,,\\
		\nabla_\mu L^\mu &=0\,, \label{eom-U}
	\end{align}
\end{subequations}
respectively, where the energy-momentum tensor is 
\begin{subequations}
	\begin{align}
		T_{\mu\nu}&=-\frac{K}{2}\Tr(L_\mu L_\nu-\frac{1}{2}g_{\mu\nu}L_\lambda L^\lambda)\,.
	\end{align}
\end{subequations}

Since we are interested in studying gravitational lensing in the presence of a nontrivial hadronic background ---hadronic lensing for short---, we must consider probe Maxwell fields propagating on a background that solves the equations of motion given in Eq.~\eqref{eom-g-U}. Since Maxwell fields couple to the pionic sector, it is necessary to introduce the $U(1)$ gauge-covariant derivative, namely,
\begin{equation}\label{gauge-covariant-derivative}
	D_\mu U=\nabla_\mu U+A_\mu U\hat{O}\,,\quad \hat{O}=U^{-1}[t_3,U]\,.
\end{equation}

For the sake of comparison, it is worth including here the action for the Abelian--Higgs model---the relativistic version of the Ginzburg--Landau free energy---to highlight the close analogies with the present case. The Lagrangian density that governs their dynamics reads~\cite{Higgs:1964pj}
\begin{equation}\label{Lag-Abelian-Higgs}
	\Lag_{\rm AH}=|D_\mu\psi|^2-\frac{1}{4}F_{\mu\nu}F^{\mu\nu}+\frac{\lambda}{2}\left(|\psi|^2-v^2\right)^2\,,
\end{equation}
where $\psi$ is a complex scalar field representing the Higgs field, $v$ is a real parameter that fixes its vacuum expectation, and $\lambda>0$ is the self-interaction coupling constant. In this case, the gauge-covariant derivative is defined as $D_\mu=\nabla_\mu - i e A_\mu$, where $e$ denotes the electric charge. Parametrizing the Higgs field in terms of its modulus and phase, i.e. $\psi=\rho e^{iS}$, the Lagrangian~\eqref{Lag-Abelian-Higgs} can be rewritten in terms of these variables as
\begin{equation}
\begin{split}
	\Lag_{\rm AH}&=(D_\mu\rho)^2+\rho^2\left(D_\mu S\right)^2-\frac{1}{4}F_{\mu\nu}F^{\mu\nu}\\
	&~~+\frac{\lambda}{2}\left(\rho^2-v^2\right)^2\,.
\end{split}
\end{equation}
with the $U(1)$ gauge-covariant derivative acting separately on the amplitude and the phase becomes
\begin{equation}
	D_\mu \rho=\nabla_\mu \rho\,,\quad D_\mu S=\nabla_\mu S - eA_\mu\,.
\end{equation}

Remarkably, a similar structure arises for the gauge-covariant derivative defined in Eq.~\eqref{gauge-covariant-derivative}. In such a case, the pionic field is parametrized as in Eq.~\eqref{U-parametrization} and the covariant derivative acting on each independent pionic degree of freedom is given by
\begin{equation}
	D_\mu \alpha=\nabla_\mu\alpha\,,\quad D_\mu\Theta=\nabla_\mu\Theta\,,\quad D_\mu\Phi=\nabla_\mu\Phi-2A_\mu\,.
\end{equation}
Thus, the scalar $\Phi(x)$ plays an analogous role to the phase of the Higgs field in the Ginzburg--Landau theory, while the remaining degrees of freedom are unaffected by the electromagnetic interaction. Substituting this parametrization into the NLSM action~\eqref{action} yields
\begin{equation}\label{I-NLSM-2}
\begin{split}
	I_{\rm NLSM}&=-\frac{K}{2}\int_\mathcal{M}\dd^4x\sqrt{-g}\Big[(\nabla\alpha)^2+\sin^2\alpha(\nabla\Theta)^2 \\
&\qquad\qquad+\sin^2\alpha\sin^2\Theta(D_\mu\Phi)^2\Big]\,.
\end{split}
\end{equation}

Since the $U(1)$ gauge connection will be treated as a probe field from hereon, its backreaction and contribution to the pionic equation will be neglected. Nevertheless, the metric and hadronic matter generate the background where the Maxwell fields are propagating, that is,
\begin{equation}\label{MaxwellEq}
	\nabla_\mu F^{\mu\nu}=J^\nu\,,  
\end{equation}
where $F_{\mu\nu}=\partial_\mu A_\nu-\partial_\nu A_\nu$ is the $U(1)$ field strength, and the electromagnetic current is given by
\begin{equation}\label{EM-current}
    J_\mu=-\frac{K}{2}\Tr(\hat{O}L_\mu)\,.
\end{equation}
Using the parametrization~\eqref{U-parametrization}, the latter reads
\begin{equation}\label{J}
    J_\mu=4K\sin^2\alpha\sin^2\Theta\left(A_\mu-\frac{1}{2}\nabla_\mu\Phi\right)\,.
\end{equation}
The metric and hadronic backgrounds encode the lens's geometric structure. This will be analyzed next in the geometric-optic limit implemented via the eikonal approximation of the Maxwell equations~\eqref{MaxwellEq}. 

\section{Eikonal approximation}\label{sec:eikonal}
To analyze how light-rays propagate in a curved spacetime in the presence of self-gravitating hadrons, let us redefine the gauge potential as $\tilde{A}_\mu=A_\mu-\frac{1}{2}\nabla_\mu \Phi$. Then, the Maxwell equations~\eqref{MaxwellEq} take a Proca-like form, i.e.,
\begin{equation}\label{eq:Maxwell-tilde}
	\nabla_\mu \tilde{F}^{\mu\nu}=\mu^2(x)\tilde{A}^\nu\,,
\end{equation}
where the effective mass term is a local function of the coordinates given by $\mu^2(x)=4K\sin^2\alpha\sin^2\Theta$. To systematically perform the eikonal approximation, we assume a WKB-type ansatz for the gauge field, consisting of a rapidly varying phase, $S(x)$, and a slowly varying amplitude, $a_\mu(x)$, that is,
\begin{equation}\label{eikonal-ansatz}
	\tilde{A}_\mu = a_\mu(x) e^{\mathrm{i} S(x)}\,,
\end{equation}
such that $|\nabla S|\gg|\nabla a|$.It is worth emphasizing that the present analysis is valid only in the low-energy regime where the photon energy remains below $\sim 200\,\mathrm{MeV}$, corresponding to the domain of applicability of the NLSM. Then, defining the wavevector $k_\mu = \nabla_\mu S$ and taking the divergence of Eq.~\eqref{eq:Maxwell-tilde} yields 
\begin{equation}
	\nabla_\mu (\mu^2\tilde{A}^\mu)=0\,.
\end{equation}
Henceforth we assume that $|\nabla S|\gg|\nabla\mu|$, namely, that the hadronic density changes slowly in comparison to the rapidly changing phase. In terms of the characteristic length scale of hadronic inhomogeneities, our approach cannot capture spatial variations occurring on scales shorter than approximately $0.5$ Fermi. Consequently, it cannot be applied to the inner quark core of neutron stars or to the very early Universe, where both matter and curvature inhomogeneities are expected to have characteristic length scales below $0.5$ Fermi. More generally, the formalism is applicable whenever the relevant physical length scales are larger than $0.5$ Fermi. This includes, for example, the study of gravitational lensing produced by compact neutron stars and black holes, as well as the propagation of electromagnetic waves through the vast majority of astrophysical plasma environments. Then, the gauge condition is reduced to the orthogonality of the wavevector and amplitude, i.e.,
\begin{equation}\label{GF}
	k_\mu a^\mu=0\,,
\end{equation}
which is the only consistent gauge-fixing condition for the Maxwell field propagating in a hadronic background.  Then, the eikonal limit of Eq.~\eqref{eq:Maxwell-tilde} reduces to the dispersion relation
\begin{equation}\label{k2}
	k_\mu k^\mu=-4K\sin^2\alpha\sin^2\Theta\,.
\end{equation}
This result shows that, in the presence of hadronic matter, light rays no longer follow null geodesics, but instead propagate along timelike trajectories, effectively behaving as massive modes. Indeed, taking the covariant derivative of Eq.~\eqref{k2}, one finds that the geodesic equation for light rays in the presence of a hadronic background is modified according to
\begin{equation}\label{new-geodesic}
    k^\mu\nabla_\mu k_\nu=-2K\nabla_\nu \left(\sin^2\alpha\sin^2\Theta\right)\,.
\end{equation}

This is an important result, as it leads to a modification of the Raychaudhuri equation~\cite{Raychaudhuri:1953yv}, which plays a central role in the proof of the singularity theorems~\cite{Hawking:1966sx,Hawking:1966vg,Hawking:1970zqf,Hawking:1973uf} (see Ref.~\cite{Kar:2006ms} for a review). This equation governs the evolution of congruences with tangent vector $k^\mu$ by describing the propagation of the expansion scalar $\theta$ along a family of curves. In particular, since these congruences no longer follow geodesics, as dictated by Eq.~\eqref{new-geodesic}, the Raychaudhuri equation acquires an additional acceleration term proportional to the divergence of the electromagnetic current, namely,
\begin{equation}\label{Ray1}
\begin{split}
	\dv{\theta}{\tau}&=-\frac{1}{3}\theta^2 -\sigma_{\mu\nu}\sigma^{\mu\nu}+\omega_{\mu\nu}\omega^{\mu\nu} - R_{\mu\nu} k^\mu k^\nu\\
    &~~-2K\Box \left(\sin^2\alpha\sin^2\Theta\right)\,,
\end{split}
\end{equation}
where $\Box\equiv g^{\mu\nu}\nabla_\mu\nabla_\nu$ denotes the d'Alembertian operator of the background metric, while $\sigma_{\mu\nu}$ and $\omega_{\mu\nu}$ are the shear and twist of $k^\mu$, respectively. The Raychaudhuri equation can also be rewritten entirely in terms of the pionic degrees of freedom. Indeed, by taking the trace of Eq.~\eqref{eom-g-U} and making use of the modified dispersion relation in Eq.~\eqref{k2}, the Ricci tensor contribution can be recast such that the modified Raychaudhuri equation becomes
\begin{align}\notag
    \dv{\theta}{\tau}&=-\frac{1}{3}\theta^2-\sigma_{\mu\nu}\sigma^{\mu\nu}+\omega_{\mu\nu}\omega^{\mu\nu}+\frac{\kappa K}{2}\Tr(L_\mu L_\nu)k^\mu k^\nu\\
    &~~-2K(\Box-2\Lambda)\sin^2\alpha\sin^2\Theta\,. \label{modified-Raychaudhuri}
\end{align}
The physical interpretation of Eq.~\eqref{modified-Raychaudhuri} is particularly transparent. The first hadronic correction, proportional to $\Tr(L_\mu L_\nu)k^\mu k^\nu$, corresponds to the standard gravitational focusing produced by the energy carried by the hadronic configuration. In contrast, the last correction originates from the spatial variation of the effective photon mass induced by the chiral medium and has no analog in vacuum GR. Consequently, the evolution of a bundle of light rays is governed by two distinct mechanisms: the spacetime curvature generated by the hadronic energy-momentum tensor and the optical response of the medium itself. The latter is analogous to light propagation through an inhomogeneous refractive medium, where refractive-index gradients alter the trajectories of light rays. In the present framework, however, the effective refractive properties are not introduced phenomenologically but are instead completely determined by the underlying hadronic fields. This makes it possible to determine, in a fully analytic manner, whether a given hadronic configuration enhances or suppresses the focusing of photon congruences.

One of the main advantages of this approach over the usual plasma lensing formalism (cf.~\cite{Bisnovatyi-Kogan:2008qbk} and references therein) is that the corrections to the Raychaudhuri equation can be computed explicitly in terms of the hadronic profiles. For instance, if the $SU(2)$-valued chiral field is known analytically, one can determine directly whether the hadronic corrections help or not the focusing of light rays. We will work out a particular case in the next section. 

\section{Hadronic lensing by superfluid pionic black holes}\label{sec:lensing-bumby-BH}

As a concrete example, we study the hadronic lensing of a black hole sourced by a nontrivial distribution of superfluid pions, recently reported in Ref.~\cite{Canfora:2026col,Canfora:2026kwj}. However, it is worth emphasizing that the present formalism is general and can be used whenever the hadrons in the background cannot be neglected.

To compute the deflection angle in the weak-field limit, we employ the Gibbons--Werner (GW) method~\cite{Gibbons:2008rj}, which provides a global geometric framework for evaluating the bending of light rays in gravitational fields. The central idea is to project the trajectories of light rays onto a two-dimensional Riemannian manifold endowed with the so-called optical metric, in which these trajectories become spatial geodesics. On this manifold, the deflection angle can be obtained by applying the Gauss--Bonnet theorem to a suitable domain of compact support bounded by the light ray and an auxiliary curve at infinity. This method was later extended to stationary and rotating spacetimes in Ref.~\cite{Werner:2012rc}, and to the case of light propagation in a cold magnetized plasma in Ref.~\cite{Crisnejo:2018uyn}. 

For a static and spherically symmetric spacetime, the GW method can be implemented in three main steps. First, the spacetime metric is parametrized as
\begin{equation}
	\dd s^2=-A(r)\dd t^2+B(r)\dd r^2+C(r)\dd\Omega^2_{(2)}\,,
\end{equation}
where $\dd\Omega^2_{(2)}=\dd\vartheta^2+\sin^2\vartheta\dd\varphi^2$ denotes the line element of the round two-sphere. Second, one considers the associated two-dimensional Riemannian manifold obtained by restricting to the equatorial plane, endowed with the optical metric
\begin{equation}
	\dd \sigma^2=\frac{n^2(r)}{A(r)}\left(B(r)\dd r^2+C(r)\dd\varphi^2\right)\,,
\end{equation}
which is conformally related to the induced metric on hypersurfaces of constant $t$ and $\vartheta=\pi/2$ of the physical spacetime. Here, $n(r)$ denotes the refractive index, which is given by
\begin{equation}
	n^2(x)=1-\frac{\omega^2_h(x )}{\omega^2(x )}\,,
\end{equation}
where $\omega(x)$ is the photon frequency measured by a static observer, while
\begin{equation}
	\omega_h^2(x )=4K\sin^2\alpha\sin^2\Theta\,,
\end{equation}
represents the effective hadronic frequency associated with the pionic distribution. Finally, applying the Gauss--Bonnet theorem, the deflection angle $\delta$ is obtained from
 \begin{equation}\label{GB-theorem}
 	\lim_{R\to \infty}\int_0^{\pi+\delta}\eval{\left(\kappa_g\dv{\sigma}{\varphi}\right)}_{C_R}\dd\varphi=\pi-\lim_{R\to \infty}\int\int_{D_R}\mathcal{K}\dd S\,,
 \end{equation}
where $D_R\subset S$ is a regular domain of the optical manifold $S$, $R$ is a constant radius to be taken to infinity, $\mathcal{K}$ is the Gaussian curvature of $D_R$, and $\kappa_g$ is the geodesic curvature of the boundary $\partial D_R$~\cite{Gibbons:2008rj}.

The line element that describes the superfluid pionic black hole of Refs.~\cite{Canfora:2026col,Canfora:2026kwj} reads
\begin{equation}\label{ds-BH}
    \dd s^2=-f(r)\dd t^2+\frac{\dd r^2}{f(r)}+r^2e^{P(x,y)}(\dd x^2+\dd y^2)\,.
\end{equation}

For the pionic degrees of freedom, we consider an ansatz that supports multi-vortex BPS configurations~\cite{Canfora:2024mkp}, i.e.,
\begin{equation}
    \alpha=2\arctan\left(e^{H(x,y)}\right)\,,\quad \Theta=\frac{\pi}{2}\,,\quad \Phi=\Phi(x,y)\,.
\end{equation}

In order to deal with analytic expressions, we restrict our analysis to the case of a pair of vortices located at the north and south poles of the horizon, with vorticities $q=1$ and $q=-1$, respectively. Nevertheless, it is worth noticing that the situation with an arbitrary number of vortices can also be addressed numerically. In the two-vortices case, an analytic solution of the pionic equation~\eqref{eom-U} is given by
\begin{equation}
	H(\rho)=\log(\rho)\,,\quad \Phi(\rho)=-\phi\,,
\end{equation}
where we have introduced the polar coordinates on the horizon, $(x,y)\mapsto(\rho\cos\phi,\rho\sin\phi)$. Focusing on a vanishing cosmological constant, the lapse function and the conformal factor $P=P(x,y)$ that solve the Einstein-NLSM field equations take the form
\begin{equation}
	f(r)=1-\frac{2m}{r}\,,\quad e^P=\frac{4(1-\kappa K)}{(1+\rho^2)^2}\,,
\end{equation}
where $m$ is an integration constant related to the mass of the solution and  $0<K<\kappa^{-1}$. Using the change of coordinate $\rho=\tan \vartheta/2$, the line element~\eqref{ds-BH} can be rewritten as
\begin{equation}
	\dd s^2=-\left(1-\frac{2m}{r}\right)\dd t^2+\frac{\dd r^2}{1-\frac{2m}{r}}+r^2(1-\kappa K)\dd\Omega^2_{(2)}\,.
\end{equation}
Notice that the lapse function is exactly that of the Schwarzschild black hole. However, it is evident that the hadronic matter induces an angular deficit. The latter will modify the optical metric and, therefore, the propagation of light rays. Additionally, although this metric has two Killing vectors, say $\partial_t$ and $\partial_\phi$, the integral curves of $k^\mu$ will generically depend on the equatorial angle, since $ k^\mu$ captures the hadronic contribution from the right-hand side of Eq.~\eqref{k2}.

In the weak-field limit, the optical metric associated with the superfluid pionic black hole is given by
\begin{equation}
	\dd \sigma^2=\frac{r(\omega_\infty^2+\omega_h^2)+2m\omega_h^2}{(r-2m)\omega_\infty^2}\left[\frac{\dd r^2}{1-\frac{2m}{r}}+r^2(1-\kappa K)\dd\varphi^2\right]\,,
\end{equation}
where $\omega_\infty$ is the photon frequency measured by a static observer at infinity, and the explicit form of the hadronic frequency in this case is $\omega_h^2=4K$. Then, a direct application of Eq.~\eqref{GB-theorem} yields the deflection angle in the weak-field limit as
\begin{equation}
	\delta=\pi\left(1-\sqrt{\frac{1}{1-\kappa K}}\right)+\frac{2m}{b}\left(1+\frac{1}{1-(\omega_h/\omega_\infty)^2}\right)\,,
\end{equation}
where $b$ denotes the impact parameter. In the limit of vanishing hadronic distribution, i.e., $K\to 0$, one recovers $\delta=4m/b$, in agreement with the standard Schwarzschild result for light deflection~\cite{Misner:1973prb,1992grle.book.....S}.

The above analytic result, in which the superfluid pions deform the black hole horizon by inducing an angular deficit, is suitable for testing our approach, since the optical metric is relatively simple. On the other hand, at least in the weak field limit, it is natural to expect that the contributions of each vortex to the light deflection should be additive, such that the total deflection can be expressed in terms of the individual vortex contributions. The opposite limit is also extremely interesting. For instance, it is worth analyzing the hadronic lensing in the strong field limit for black holes supported by hadronic matter, such as the superfluid pionic black hole discussed in Ref.~\cite{Canfora:2013osa}, following the pioneering references~\cite{Bozza:2001xd,Bozza:2002zj,Bozza:2002af,Bozza:2007gt,Bozza:2006sn}. We will come back to these questions in the future.

\section{Conclusions}\label{sec:conclusions}
We introduced a novel analytic approach to incorporate hadronic effects into gravitational lensing analyses. These effects appear in backgrounds that include not only a gravitational field but also a nontrivial distribution of hadrons. Our analysis is relevant in astrophysics and cosmology when hadrons cannot be neglected. Photon-hadron interactions are very similar to photon-Cooper-pair interactions, in which photons acquire an effective mass term that may depend on the coordinates if the condensate is not homogeneous. In such a case, photons do not follow null geodesics. The hadronic degrees of freedom can be described by the nonlinear sigma model, which corresponds to the leading-order term of chiral perturbation theory in the pion decay constant, minimally coupled to Maxwell's gauge theory. 

The modified Raychaudhuri equation, which includes hadronic corrections, was derived, along with the corresponding dispersion relation for photons propagating in a hadronic medium. The present results are fully consistent with the theory of gravitational lensing by a plasma medium. However, it has the advantage that the plasma transport properties, such as the refractive index, can be expressed analytically in terms of the hadronic matter distribution. This allows one to determine whether the $SU(2)$-valued chiral field helps focus light rays. As an example, we have computed the weak-limit deflection angle for an analytic black hole in General Relativity sourced by superfluid pionic vortices. However, our results are general and apply to any gravitational-hadronic background. 

It is worth emphasizing that the theoretical framework developed in this work is more general than the explicit black hole example considered in Sec.~\ref{sec:lensing-bumby-BH}. The Maxwell equations, which include photon-hadron interactions, the effective dispersion relation, and the generalized Raychaudhuri equation, are model-independent consequences at leading order in chiral perturbation theory, together with the eikonal approximation. In contrast, the explicit deflection angle derived in Sec.~\ref{sec:lensing-bumby-BH} depends on the particular superfluid pionic black hole solution adopted as an illustrative example.

Interesting questions remain open. For instance, it would be relevant to study superfluid multi-vortex pionic black holes to examine the effect on light deflection in both the weak- and strong-field limits. Additionally, studying the Maxwell equations without the eikonal approximation is worth exploring. For instance, if the photons propagate in a nonhomogeneous hadronic medium, our approach could be used to study hadronic birefringence and mode splitting. Finally, analyzing the backreaction of photons coupled to superfluid pionic vortices could open interesting avenues for studying black holes with charged nonlinear matter fields. We leave these and other questions for future investigations.

\begin{acknowledgments}
    The authors thank Jorge Cuadra, Astor Sandoval, Leonardo Sanhueza, and Jorge Zanelli for discussions. This work is partially funded by the Agencia Nacional de Investigación y Desarrollo (ANID) through Fondecyt Regular grants No. 1240043, 1240048, 1252053, 1251523, 1261016, as well as by ANID Proyecto de Exploración 1325001.
\end{acknowledgments}

\bibliography{References}

\end{document}